\documentclass[twocolumn,citeautoscript,superscriptaddress,8pt]{revtex4-2}

\usepackage{hyperref}
\usepackage{graphicx}
\usepackage{gensymb}
\usepackage{color}
\usepackage[dvipsnames]{xcolor}
\usepackage{float}
\usepackage{upgreek}
\usepackage{bm}
\usepackage{amsmath,amssymb}

\begin{document}

\title{Quantum microscopy with van der Waals heterostructures}

\author{A.~J.~Healey} 
\thanks{These authors contributed equally to this work.}
\affiliation{School of Physics, University of Melbourne, VIC 3010, Australia}
\affiliation{Centre for Quantum Computation and Communication Technology, School of Physics, University of Melbourne, VIC 3010, Australia}

\author{S.~C.~Scholten} 
\thanks{These authors contributed equally to this work.}
\affiliation{School of Physics, University of Melbourne, VIC 3010, Australia}

\author{T.~Yang} 
\thanks{These authors contributed equally to this work.}
\affiliation{School of Mathematical and Physical Sciences, University of Technology Sydney, Ultimo, NSW 2007, Australia}
\affiliation{ARC Centre of Excellence for Transformative Meta-Optical Systems, University of Technology Sydney, Ultimo, NSW 2007, Australia}

\author{J.~A.~Scott} 
\affiliation{School of Mathematical and Physical Sciences, University of Technology Sydney, Ultimo, NSW 2007, Australia}
\affiliation{ARC Centre of Excellence for Transformative Meta-Optical Systems, University of Technology Sydney, Ultimo, NSW 2007, Australia}

\author{G.~J.~Abrahams} 
\affiliation{School of Physics, University of Melbourne, VIC 3010, Australia}

\author{I.~O.~Robertson} 
\affiliation{School of Physics, University of Melbourne, VIC 3010, Australia}
\affiliation{School of Science, RMIT University, Melbourne, VIC 3001, Australia}

\author{X.~F.~Hou} 
\affiliation{School of Physical Science and Technology, ShanghaiTech University, Shanghai 201210, China}

\author{Y.~F.~Guo} 
\affiliation{School of Physical Science and Technology, ShanghaiTech University, Shanghai 201210, China}
\affiliation{ShanghaiTech Laboratory for Topological Physics, ShanghaiTech University, Shanghai 201210, China}

\author{S.~Rahman} 
\affiliation{School of Engineering, College of Engineering \& Computer Science, Australian National University, Canberra, ACT 2601, Australia}

\author{Y.~Lu} 
\affiliation{School of Engineering, College of Engineering \& Computer Science, Australian National University, Canberra, ACT 2601, Australia}
\affiliation{Centre for Quantum Computation and Communication Technology, Australian National University, Canberra, ACT 2601, Australia}

\author{M.~Kianinia} 
\affiliation{School of Mathematical and Physical Sciences, University of Technology Sydney, Ultimo, NSW 2007, Australia}
\affiliation{ARC Centre of Excellence for Transformative Meta-Optical Systems, University of Technology Sydney, Ultimo, NSW 2007, Australia}

\author{I.~Aharonovich}
\email{igor.aharonovich@uts.edu.au}
\affiliation{School of Mathematical and Physical Sciences, University of Technology Sydney, Ultimo, NSW 2007, Australia}
\affiliation{ARC Centre of Excellence for Transformative Meta-Optical Systems, University of Technology Sydney, Ultimo, NSW 2007, Australia}

\author{J.-P.~Tetienne}
\email{jean-philippe.tetienne@rmit.edu.au}
\affiliation{School of Science, RMIT University, Melbourne, VIC 3001, Australia}

\begin{abstract} 

Quantum microscopes based on solid-state spin quantum sensors have recently emerged as powerful tools for probing material properties and physical processes in regimes not accessible to classical sensors, especially on the nanoscale \cite{Rondin2014,Casola2018,Levine2019,Scholten2021}. Such microscopes have already found utility in a variety of problems, from imaging magnetism and charge transport in nanoscale devices \cite{Tetienne2014,Gross2017,Tetienne2017a,Thiel2019,Ku2019,Broadway2020}, to mapping remanent magnetic fields from ancient rocks and biological organisms \cite{LeSage2013,Fu2014,deGille2021}. However, applications of quantum microscopes have so far relied on sensors hosted in a rigid, three-dimensional crystal, typically diamond, which limits their ability to closely interact with the sample under study. Here we demonstrate a versatile and robust quantum microscope using quantum sensors embedded within a thin layer of a van der Waals (vdW) material, hexagonal boron nitride (hBN). To showcase the capabilities of this platform, we assemble several active vdW heterostructures, with an hBN layer acting as the quantum sensor. We demonstrate time-resolved, simultaneous temperature and magnetic imaging near the Curie temperature of a vdW ferromagnet as well as apply this unique microscope to map out charge currents and Joule heating in graphene. By enabling intimate proximity between sensor and sample, potentially down to a single atomic layer, the hBN quantum sensor represents a paradigm shift for nanoscale quantum sensing and microscopy. Moreover, given the ubiquitous use of hBN in modern materials and condensed matter physics research \cite{Dean2010,Chen2020}, we expect our technique to find rapid and broad adoption in these fields, further motivated by the prospect of performing in-situ chemical analysis and noise spectroscopy using advanced quantum sensing protocols \cite{Glenn2018,Kolkowitz2015,Du2017,Simpson2017}.

\end{abstract}

\maketitle 

\clearpage

Quantum sensing has garnered vast attention over the last decade, thanks to the advent of robust solid-state spin defects, which enable a range of practical applications across condensed matter physics \cite{Casola2018}, geoscience \cite{Glenn2018}, and biology \cite{Schirhagl2014}. To date, the leading spin system for quantum sensing has been the nitrogen-vacancy centre in diamond, although spin defects have been found and studied in other solid-state materials such as silicon carbide and silicon \cite{Wolfowicz2021}. The bulk, three-dimensional (3D) nature of these host materials provides natural protection for the embedded quantum sensors, but it also makes their precise interfacing with external objects difficult, and hampers integration of quantum sensors into complex devices. Moreover, for near-surface defects the perturbation to the 3D structure leads to worsened performance and stability issues. Quantum microscopy, which employs quantum sensors to form a spatially-resolved map of a physical quantity such as the magnetic field, is particularly affected by these limitations since both spatial resolution and sensitivity directly depend on the standoff between sensor and sample \cite{Rondin2014,Casola2018}.

\begin{figure*}[t!]
\centering
\includegraphics[width=0.8\textwidth]{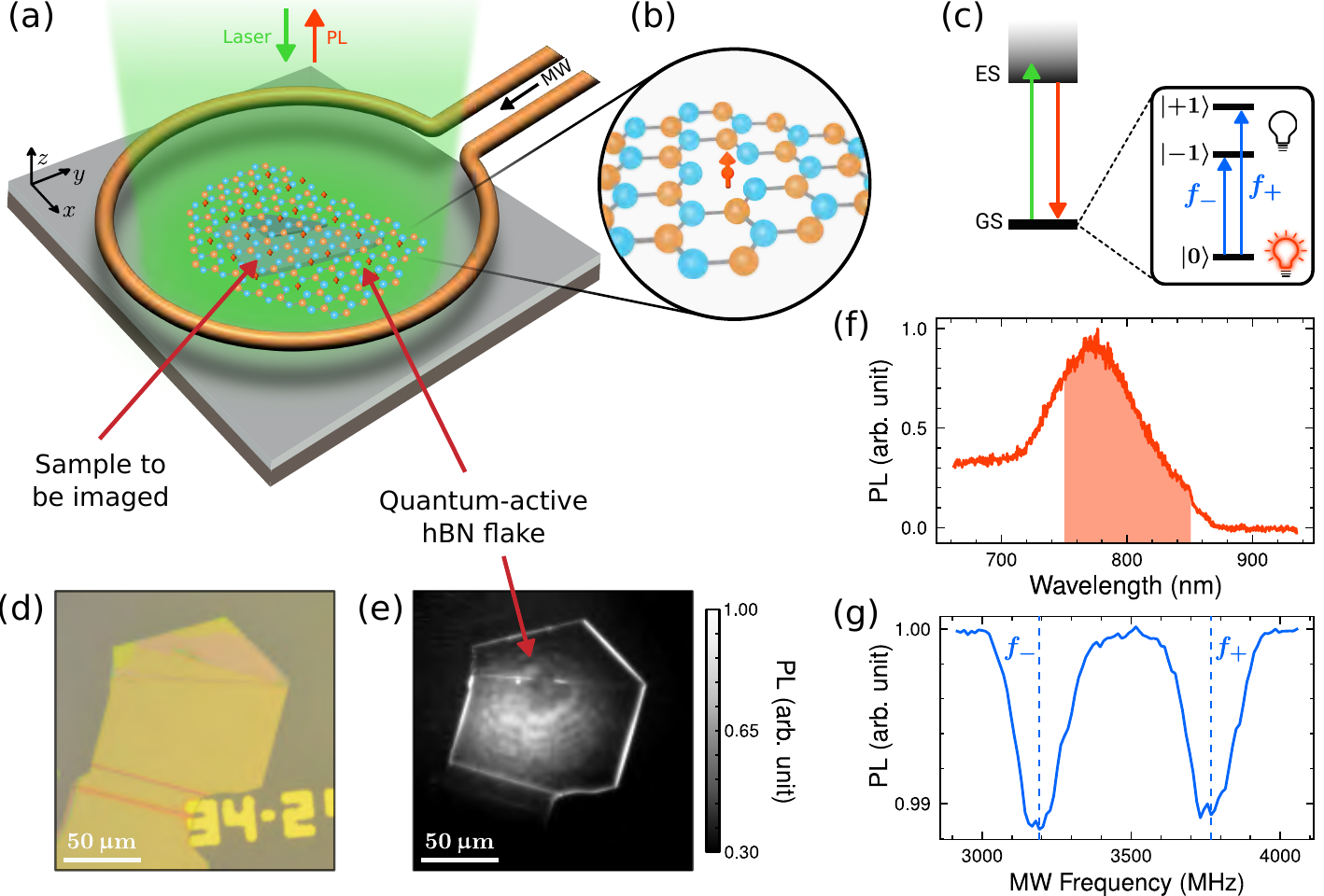}
\caption{{\bf Quantum microscopy with spin defects in hBN}. (a) Conceptual schematic of the experiment. An hBN flake containing $V_{\rm B}^-$ spin defects (red arrows) is placed on the sample under study. The defects are excited by a laser and a microwave (MW) field, and their photoluminescence (PL) is imaged with a camera to enable spatially resolved optically detected magnetic resonance (ODMR). (b) Atomic representation of the $V_{\rm B}^-$ defect (nitrogen atoms in blue, boron in orange). (c) Simplified energy level diagram of the $V_{\rm B}^-$ defect showing the optical transitions between ground state (GS) and excited state (ES), and the MW transitions between GS spin sub-levels $|0,\pm1\rangle$. (d) Optical image of a typical hBN flake. (e) Corresponding PL image under widefield laser illumination ($\lambda=532$~nm). (f) PL spectrum of the flake in (e). The shading indicates the emission band used for ODMR (750-850~nm). (g) ODMR spectrum under a bias magnetic field $B_0\approx10$~mT parallel to the $z$ axis.}
\label{fig1}
\end{figure*}

A potential solution for this increasingly important challenge is the utilisation of vdW crystals that could be exfoliated down to a few atomic layers. Such a system would enable materials and devices located just a few atoms away to be probed, bringing quantum sensing and microscopy into a hitherto unexplored regime. Furthermore, the seamless integration of vdW materials into multifunctional heterostructures \cite{Novoselov2016} would open new prospects for in-situ quantum sensing, with potential for widespread use in a range of devices and scenarios. Indeed, a long-standing goal has been to identify a spin defect housed in a vdW material that would be suitable for quantum microscopy \cite{Liu2019,Bassett2019,Ping2021,Gottscholl2020,Chejanovsky2021,Stern2021}. 

Here we present the first demonstration of a quantum microscope using a vdW material as the active sensor. Specifically, we employ spin defects in hexagonal boron nitride (hBN), a wide bandgap vdW crystal. Importantly, hBN is already routinely utilised in vdW heterostructures as an electrical insulator~\cite{Novoselov2016} and is stable in air even when exfoliated down to a few monolayers, which is vital for a versatile quantum sensor. 
The spin defect is the negatively charged boron vacancy centre, ($V_{\rm B}^-$), a spin-1 system that exhibits a robust optically detected magnetic resonance (ODMR) response over a wide range of temperatures up to 600 K \cite{Gottscholl2021,Gottscholl2021b,Liu2021,Gao2021}. 

Our concept for the vdW quantum microscope is depicted in Fig.~\ref{fig1}a. A thin flake of hBN containing a high density of $V_{\rm B}^-$ defects (Fig.~\ref{fig1}b) is placed directly on the sample to be imaged. A widefield optical microscope is used to excite the $V_{\rm B}^-$ defects and image their spin-dependent photoluminescence (PL) with a camera. A nearby microwave (MW) antenna allows the electron spin transitions (with frequencies $f_\pm$, see Fig.~\ref{fig1}c) to be interrogated via ODMR spectroscopy. These transition frequencies depend on the local environment (e.g., magnetic field, temperature, pressure~\cite{Gottscholl2021b}), revealing insights into the sample's properties. We prepared large ($\sim100~\mu$m laterally) hBN flakes with a 10-100 nm thickness by exfoliation and subsequently created the $V_{\rm B}^-$ defects by ion irradiation. The flakes are then deterministically transferred onto a sample of interest for imaging. An example flake is shown in Fig.~\ref{fig1}d. The corresponding PL image under widefield laser illumination (Fig.~\ref{fig1}e) reveals a relatively uniform emission in the centre of the flake, with a PL spectrum (Fig.~\ref{fig1}f) characteristic of the $V_{\rm B}^-$ defect. An example ODMR spectrum is shown in Fig.~\ref{fig1}g, revealing the two spin resonances used for quantum sensing.

\begin{figure}[t]
\centering
\includegraphics[width=0.45\textwidth]{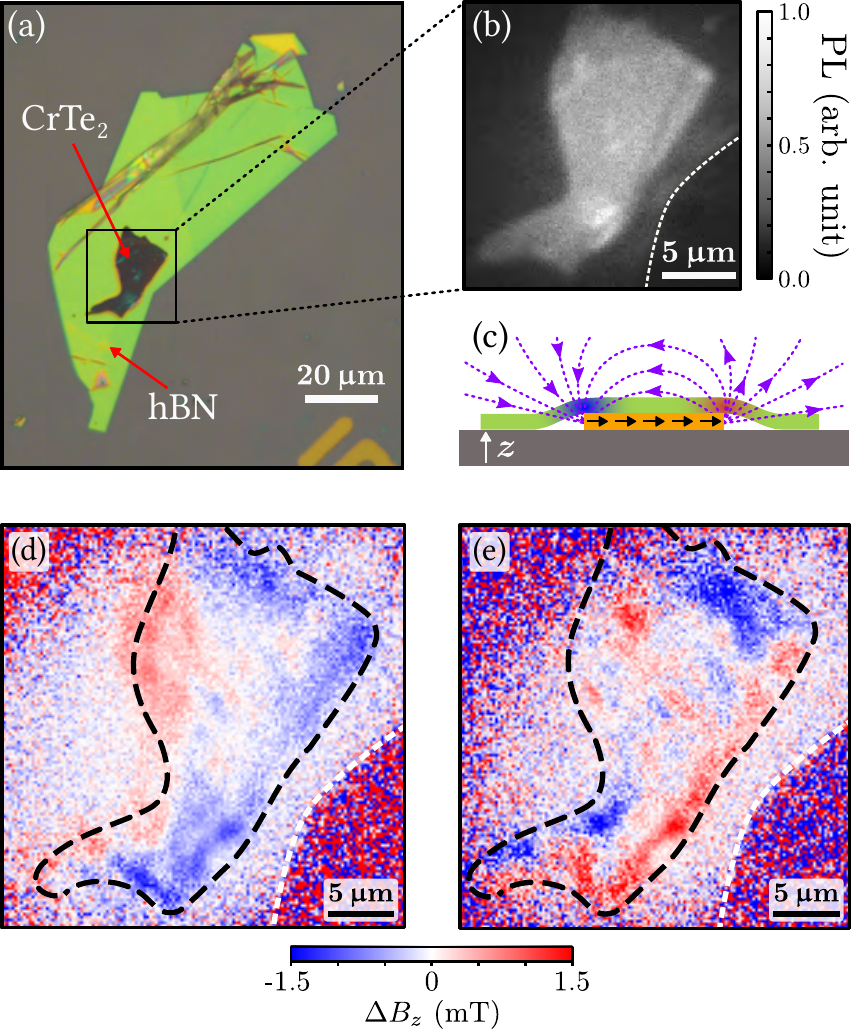}
\caption{{\bf Magnetic imaging of a CrTe$_2$ ferromagnet}. (a) Optical image of a CrTe$_2$ flake (dark region) covered by an hBN flake (green). (b) PL image centered on the CrTe$_2$ flake. White dashed line indicates the edge of the hBN flake. (c) Sketch of the stray magnetic field (purple lines) produced by a flake with in-plane magnetisation (depicted by black arrows). The hBN quantum sensor measures the $z$ projection of the field, as depicted by the red and blue shading. (d) Stray magnetic field map ($\Delta B_z$) of the CrTe$_2$ flake. Dashed black line indicates the flake's contour. (e) $\Delta B_z$ map obtained after inverting the direction of the bias magnetic field, causing partial magnetisation reversal.}
\label{fig2}
\end{figure} 

To demonstrate the performance of hBN for quantum microscopy, we first image the stray magnetic field of a vdW ferromagnet, CrTe$_2$. Namely, we prepared a thin CrTe$_2$ flake (70 nm) on a SiO$_2$/Si substrate, and covered it with an active, 40-nm-thick hBN flake (see optical image in Fig.~\ref{fig2}a). The corresponding PL image (Fig.~\ref{fig2}b) reveals an increased PL above the CrTe$_2$ flake due to reflection from the metal \cite{Gao2021}. By fitting the ODMR spectrum for each pixel, a map of the magnetic field projected in the hBN flake ($z$ component, see Fig.~\ref{fig2}c) can be obtained via the Zeeman relation, $B_z=(f_+-f_-)/2\gamma_e$, where $\gamma_e=28$~GHz/T is the electron gyromagnetic ratio~\cite{Gottscholl2020}. Subtracting the applied bias magnetic field ($B_0\approx10$~mT) then yields the stray field emanating from the CrTe$_2$ flake, $\Delta B_z=B_z-B_0$ (Fig.~\ref{fig2}d). The observed pattern is characteristic of an in-plane magnetisation, and the amplitude (up to $\pm1.5$~mT) is consistent with a spontaneous magnetisation $M_s\approx 50$~kA/m, in agreement with expectation (see SI, Sec. XII). By reversing the direction of the bias field, we observe a partial magnetisation reversal (Fig.~\ref{fig2}e), indicating a coercive field of comparable magnitude to the 10 mT bias field~\cite{Purbawati2020}.

\begin{figure}[b]
\centering
\includegraphics[width=0.5\textwidth]{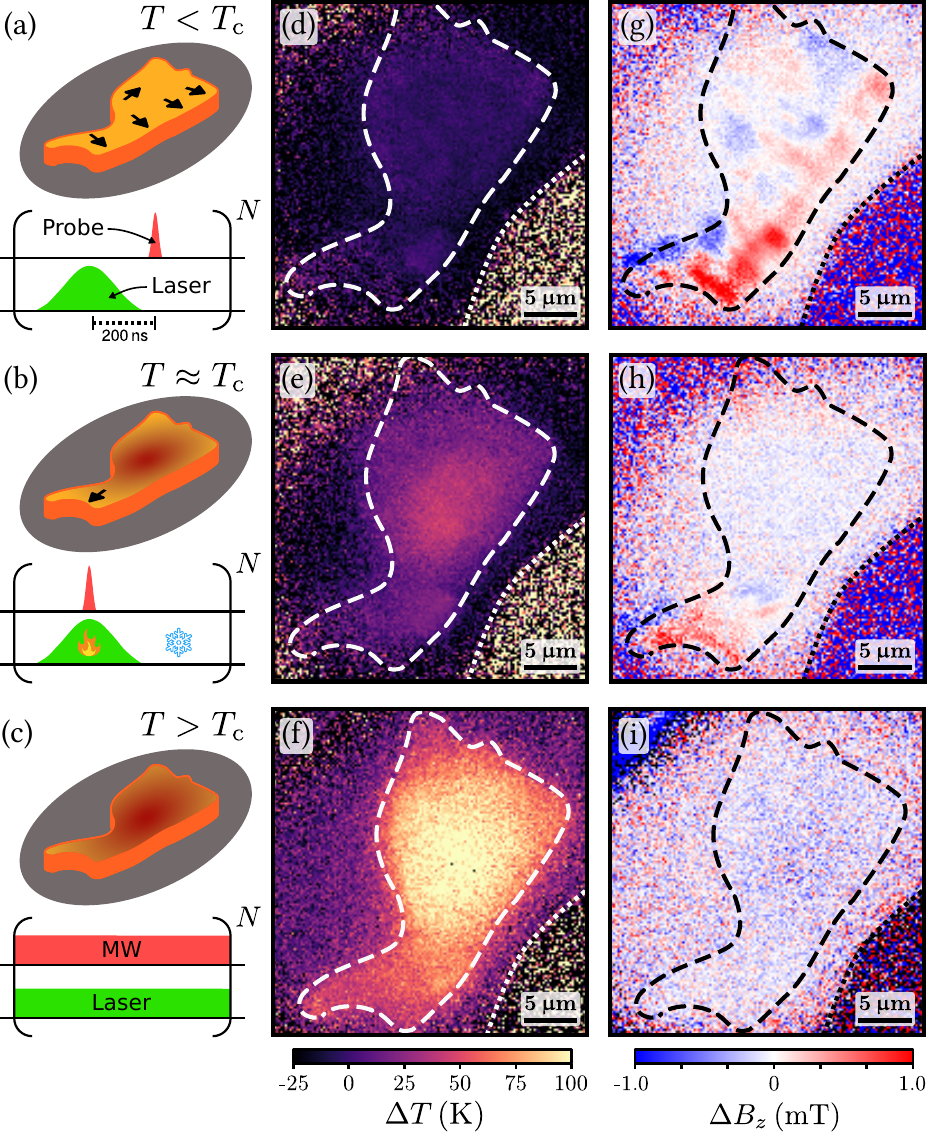} 
\caption{{\bf Time-resolved simultaneous temperature and magnetic imaging}. (a-c) Schematics depicting the pulse sequences used (green: laser pulse; red: MW pulse) and their effect on the CrTe$_2$ flake's magnetisation (black arrows). In (a,b), the 40-ns MW probe pulse is applied (a) 200 ns after and (b) during the peak of the laser pulse. In (c), both laser and MW are applied continuously. (d-f)  
Temperature maps ($\Delta T$) of the CrTe$_2$ flake taken under the conditions depicted in (a-c) respectively. (g-i) Corresponding stray magnetic field maps ($\Delta B_z$).}
\label{fig3}
\end{figure}

We next demonstrate temperature imaging, exploiting the fact that the zero-field splitting parameter of the $V_{\rm B}^-$ defects, $D=(f_++f_-)/2$, is directly related to the temperature of the hBN crystal~\cite{Gottscholl2021b,Liu2021}. Near room temperature, we can use the linear approximation, $D\approx D_0-\alpha \Delta T$ where $D_0=3480$~MHz, $\alpha=0.70$~MHz/K and $\Delta T=T-300$~K. As the hBN flake is in direct contact with the sample and is not connected to a thermal reservoir other than the sample's substrate, it will act as an accurate, minimally invasive reporter of the sample's temperature. Using the same hBN/CrTe$_2$ heterostructure, we mapped $\Delta T$ under different laser illumination conditions (depicted in Fig.~\ref{fig3}a-c), as laser absorption by the CrTe$_2$ is expected to cause heating. 

Figure~\ref{fig3}d shows the $\Delta T$ map obtained with a pulsed ODMR protocol where the MW pulse (which encodes $\Delta T$ into the PL) is delayed by 200\,ns with respect to the laser pulse, resulting in a modest temperature increase in the CrTe$_2$ flake of $\approx 15$\,K above the background value. However, when the MW pulse is brought forward to probe the temperature at the peak of the laser pulse, a larger increase up to $\Delta T\approx 50$~K is observed (Fig.~\ref{fig3}e). Moreover, the temperature profile correlates with the laser spot profile (i.e. maximum near the centre of the image), in contrast with Fig.~\ref{fig3}d where the temperature appears more uniform as the CrTe$_2$ flake has thermalised. Using a continuous-wave ODMR protocol with no cooling time, the temperature is even higher (Fig.~\ref{fig3}f) with a maximum of $\Delta T\approx 100$~K near the centre, and a measurable bleeding into the substrate ($\Delta T\approx 30$~K just outside the CrTe$_2$ flake). 

These temperature maps can be correlated with the magnetic field maps obtained from the same quantum microscopy data (Fig.~\ref{fig3}g-i). While a complete magnetic field pattern is recovered after the cooling time (Fig.~\ref{fig3}g, similar conditions to Fig.~\ref{fig2}e), the magnetisation appears partly quenched during the laser pulse (Fig.~\ref{fig3}h) and the magnetic signal completely disappears under continuous illumination (Fig.~\ref{fig3}i). These results are consistent with the Curie temperature of CrTe$_2$ which is only slightly above room temperature ($T_c\approx315$~K~\cite{Purbawati2020}). 

Importantly, these results illustrate the unique ability to perform time-resolved -- in a stroboscopic manner, via the tunable delay of the probe MW pulse -- simultaneous temperature and magnetic field imaging. The temporal resolution is given by the duration of the MW pulse, 40~ns in Fig.~\ref{fig3}, which in principle can be reduced to a few nanoseconds~\cite{Fuchs2009}. We stress that the ability to accurately report the sample's temperature is a direct consequence of the thin vdW nature of the sensor, contrasting with bulk (e.g. diamond) based quantum sensors which are naturally good thermal reservoirs and therefore ill-suited to temperature imaging. 

\begin{figure*}[t]
\includegraphics[width=0.8\textwidth]{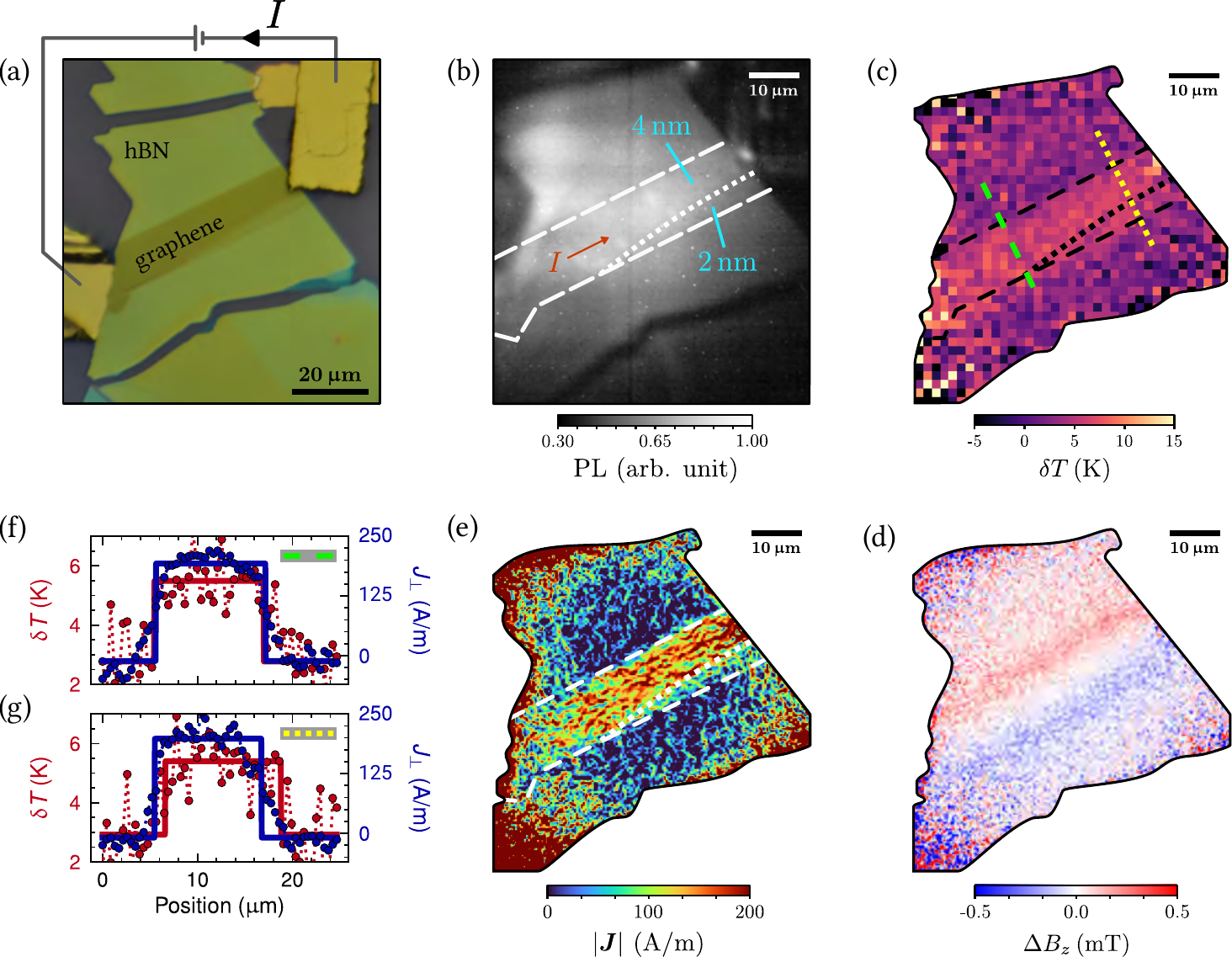}
\caption{{\bf Imaging Joule heating and current flow in a graphene device.} (a) Optical image of the hBN/graphene device and its electrical connections. (b) PL image, with the edges of the few-layer graphene ribbon shown as dashed white lines. The dotted line indicates a thickness change from 4 nm (above the line) to 2 nm (below). (c) Map of the temperature change $\delta T$ induced by a constant current $I=3$~mA, referenced to the no-current case. (d) Stray magnetic field map ($\Delta B_z$) under the same $I=3$~mA current. (e) Current density map (norm $|{\bf J}|$) reconstructed from (d). (f,g) Linecuts of $\delta T$ (red dots) and $J_\perp$ (blue dots) taken along two different lines drawn in (c): (f) across a region of the ribbon with uniform thickness; (g) across a region that has a 2-nm-thick section. $J_\perp$ is the component of the current density vector perpendicular to the linecut direction. Solid lines are rectangular function fits.}
\label{fig4}
\end{figure*}

As a final proof of the versatility of the vdW quantum microscope, we image the temperature and current density of an operating two-terminal graphene device. The device consists of a few-layer graphene ribbon (4 nm thick except for a small 2-nm region) covered with a 70-nm hBN flake and connected to gold electrodes (Fig.~\ref{fig4}a). The PL image reveals uniform emission from the hBN flake with little contrast from the graphene (Fig.~\ref{fig4}b). We recorded temperature and magnetic field maps under a constant current $I=3$~mA flowing through the device, and normalised the data to the no-current case (see full data in SI, Sec. XII). First, we observe a temperature increase of $\delta T\approx 6$~K in the graphene ribbon (Fig.~\ref{fig4}c), as a consequence of Joule heating. Here $\delta T$ is the change due to the current alone, subtracting the effect of laser heating (see SI, Sec. XIII). Second, a clear magnetic field pattern emerges near the ribbon (Fig.~\ref{fig4}d), which is characteristic of a current-induced (Oersted) field. Using a Fourier-domain inversion method~\cite{Tetienne2017a,Broadway2020a}, we reconstructed the 2D current density map in the device (Fig.~\ref{fig4}e). Interestingly, linecuts across the ribbon taken in two different locations (Fig.~\ref{fig4}f,g) reveal that while the current density is mainly confined to the thicker section of the ribbon, the temperature increase extends to the thinner section (see Fig.~\ref{fig4}g where the $\delta T$ increase spreads further to the right than the current). This observation is a direct illustration of the high thermal conductivity of graphene. These results illustrate the enormous potential of our technique for spatially resolved, combined thermal and transport studies of graphene and other 2D material systems.

We now discuss the current performance and expected future developments of the vdW quantum microscope. The spatial resolution is limited by diffraction, about $1~\mu$m in this work, but in principle could be reduced down to $\approx10$~nm using super-resolution nanoscopy techniques~\cite{Comtet2019,Chen2019}. The sensitivity in our images (estimated from the pixel-to-pixel noise) was about $100~\mu$T for magnetic field and 5~K for temperature, for a $220\times220$~nm$^2$ pixel after several hours of integration. We anticipate that a two order of magnitude improvement in sensitivity can be achieved through optimisation of the materials and experimental set-up (see SI, Sec. VIII). Further sensitivity improvements are expected for the measurement of signals oscillating (coherently or randomly) at kHz-MHz frequencies, via the implementation of quantum lock-in protocols~\cite{Staudacher2013}, which should also enable in-situ chemical analysis of the sample~\cite{Glenn2018}. Higher frequency (GHz) signals can be accessed via spin relaxometry ($T_1$) techniques, enabling charge noise and spin wave spectroscopy, and ion sensing \cite{Casola2018,Kolkowitz2015,Du2017,Simpson2017}. 

Crucially, the hBN layer can be readily thinned down to a few monolayers and is able to conform to the sample's topography, allowing atomic proximity over large imaging areas. This attribute is what makes vdW quantum sensors truly unique compared to their 3D counterparts, along with the demonstrated multi-modal imaging capability. These key enabling features open exciting new opportunities both in the physical sciences, e.g. for real-space investigations of emerging phenomena in twisted vdW heterostructures~\cite{Cao2018,Song2021}, and in the life sciences where long standing challenges such as mapping of ion channels and proteins may become possible~\cite{Hall2010,Abobeih2019}. Finally, we stress that the vdW quantum microscope is relatively easy to implement, relying on readily available vdW materials and routine exfoliation and transfer methods. As such, we anticipate that the technique will be rapidly adopted to tackle a broad range of open questions across a myriad of scientific fields.

\begin{acknowledgments}
This work was supported by the Australian Research Council (ARC) through grants CE170100012, DP190101506, FT200100073 and CE200100010, and the Asian Office of Aerospace Research \& Development (FA2386-20-1-4014). The authors thank the Australian Nanofabrication Facilities at the UTS OptoFab node. A.J.H. is supported by an Australian Government Research Training Program Scholarship. S.C.S gratefully acknowledges the support of an Ernst and Grace Matthaei scholarship. Y.F.G. acknowledges the support by the National Natural Science Foundation of China (Grant Nos. 92065201 and 11874264).
\end{acknowledgments}







\bibliographystyle{naturemag} 
\bibliography{library}

\end{document}